\documentclass{aa}

\usepackage{graphicx}
\usepackage{txfonts}
\usepackage{natbib}
\usepackage{ulem}
\begin{document}
\title{On the validity of the $630$ nm Fe\,{\sc i} lines for the magnetometry of
the internetwork quiet Sun}

\author{M. J. Mart\' inez Gonz\'alez, M. Collados \and B. Ruiz Cobo}
\institute{Instituto de Astrof\'{\i}sica de Canarias, 38205, La Laguna, Tenerife, Spain}

\date{Received <date>; accepted <date>}


\abstract
{}
{The purpose of this work is to analyze the reliability of the magnetic field strengths inferred from 
the $630$ nm pair of Fe\,{\sc i} lines at internetwork quiet Sun regions .}
{Some numerical experiments have been performed that demonstrate
the inability of these lines to recover the magnetic field strength
in such low flux solar regions.}
{It is shown how different model
atmospheres, with magnetic field strengths ranging from few hundred Gauss
to kiloGauss, give rise to Stokes profiles that can not be distinguished.
The reasons for this degeneracy are discussed.}
{}


\authorrunning{M. J. Mart\' inez Gonz\'alez et al.}
\titlerunning{Validity of $630$ nm QS magnetometry}
\maketitle


\section{Introduction}

The magnetometry of quiet Sun regions is strongly restrained by
important difficulties. Basically, the very small circular polarization
signals and the fact that structures in those regions are unresolved
make the interpretation with standard techniques complicated.
It is not a surprise that different works based on data
taken with different instruments (spectroscopic or imaging polarimetry),
various wavelengths (near infrared or visible spectral lines) and
analysis procedures (line ratio, inversions following ME, LTE or MISMA schemes) lead to a diversity
in the results larger than the expected from observational uncertainties.

Most works studying quiet Sun regions in the visible part of the spectrum
have used the Fe\,{\sc i} pair of lines located at 630 nm. Among them, one
can cite, for instance, \cite{hector_02}, \cite{ita_jorge_03}, \cite{hector_valentin_04}, 
and \cite{lites_04}. Simple methods, such as the line ratio technique, and also more 
sophisticated inversion procedures, have been used to infer magnetic field 
strengths in network and internetwork regions. But, until now, no deep 
study of the physical properties of this pair of lines has been performed 
to support or to refuse their validity for the analysis of the quiet
Sun magnetism in low flux regions (such as the internetwork). \cite{luis_03} 
already pointed out some restrictions of the \hbox{630 nm} lines
due to uncertainties introduced by noise.

Results from the near infrared Fe\,{\sc i} lines at 1.56 microns 
usually lead to the conclusion that most of the magnetic fields in the
internetwork are in the form of weak fields \citep[see][]{lin_95, lin_99, khomenko_03}. This result is
confirmed by \cite{tip_polis_06} but
contradicted by \cite{jorge_ita_03} and \cite{ita_05}. These last
three works analyze simultaneous \hbox{$1.5$ $\mathrm{\mu}$m} and $630$ nm observations.

As explained below, the often used Fe\,{\sc i} pair of lines located at
630 nm suffer from a major drawback: their different formation heights 
makes them sensitive to different parts of the atmosphere. As a consequence,
all those results obtained using analysis techniques that ignore this
fact shoud be regarded with care.

The structure of this paper is as follows: first, we demonstrate that the 
Fe\,{\sc i} 630 nm line pair have their maximum sensitivity at different
layers in the quiet Sun atmosphere. Second, we show how low
amplitude circular polarization signals (like those present in the internetwork quiet Sun)
can be retrieved using model atmospheres with weak or strong magnetic fields,
provided one allows other atmospheric parameters to compensate for the
magnetic field effect. Also, we show that the amplitude ratio of the
circular polarization signals is not a good indicator of the magnetic field
strength. Finally, different PDFs resulting from the inversion of real data
are presented.


\section{Different line formation heights}

Response Functions (RFs) \citep{basilio_94,jorge_basilio_96}
are suitable indicators of line formation height. The maximum of a RF to 
the perturbation of a parameter provides an estimate of the height in the atmosphere
where a given spectral line is most sensitive to such parameter.

The RFs to several physical magnitudes have been calculated for the Fe\,{\sc i}
lines at 630.1 nm and at 630.2 nm. For the calculations, an HSRA model atmosphere
\citep{hsra} has been used as representative of the quiet Sun. Several different 
configurations with constant, vertical magnetic field with values ranging from 
$0$ to $1000$ G and macroturbulent values between $0$ and $2$ km/s have been 
used. No bulk velocity has been included. A constant with height microturbulent 
velocity of $0.6$ km/s has been added to the model atmosphere.

\begin{figure}[!t]
\includegraphics[width=8cm]{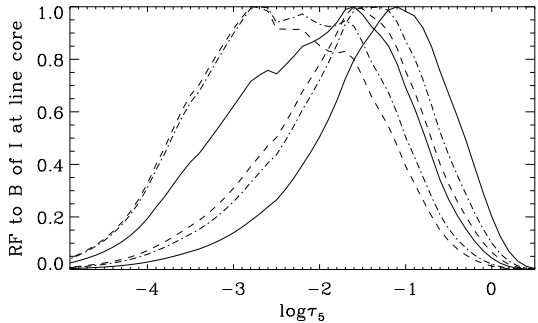}
\caption{Response Function to the magnetic field of the intensity line core of the $630.1$ nm
and $630.2$ nm lines using an HSRA quiet Sun model atmosphere \citep{hsra}
with different constant vertical magnetic fields. $\tau_5$ is the continuum optical depth at 
$500$ nm. Thick (thin) lines are the RF for the $630.2$ nm ($630.1$ nm) line. Solid lines are the RF when the magnetic field is set to $1000$ G, dashed-dotted lines correspond to $500$ G and dashed lines to $200$ G. The macroturbulent velocity has been fixed to $1$ km/s.}
\label{alturas_formacionB}
\end{figure}

Figure \ref{alturas_formacionB} shows some RFs to the magnetic field strength
of the line core intensity. In all of the cases, for a given magnetic
field strength, the 630.1 line has the maximum of the RF at higher layers than
the 630.2 line, being the difference larger for lower magnetic field strengths.

\begin{figure}[!t]
\includegraphics[width=8cm]{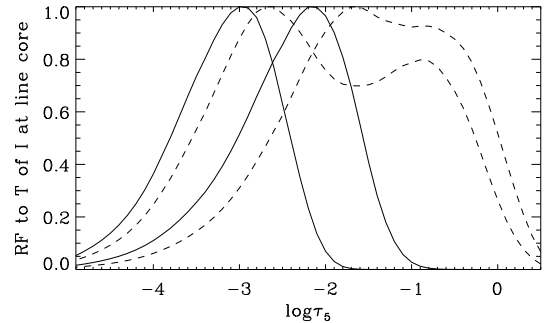}
\caption{Response Function to the temperature of the intensity line core of the $630.1$ nm
and $630.2$ nm lines using an HSRA quiet Sun model atmosphere \citep{hsra}
with two different values for the macroturbulent velocity. Thick lines correspond to the $630.2$ nm line and the thin lines to the $630.1$ nm ones. The solid lines are the RF when the macroturbulent velocity has a value of $0$ km/s while dashed lines correspond to a value of $2$ km/s. No magnetic field
has been introduced.}
\label{alturas_formacionT}
\end{figure}

Figure \ref{alturas_formacionT} presents the RFs to temperature of the line
core intensity, for the $B=0$ model atmosphere and with the extreme values
of the macroturbulent velocity. This parameter works as a convolution with
the Stokes profiles. As a consequence, a mixing of the information
at each wavelength occurs. The higher the macroturbulence velocity, the larger the
information blurring, and the lower the height corresponding to
the maximum of the RF is. As in the
previous case of the RF to magnetic field strength, now, for all macroturbulent
velocity values, both lines are sensitive to different temperature layers
in the atmosphere, the 630.1 nm line tracing higher layers than the $630.2$ nm one. This result
is in agreement with that of \cite{shchukina_01}, where these authors studied the formation 
height of many Fe\,{\sc i} spectral lines, including NLTE effects.

\section{LTE inversions}

Inversion methods are powerful techniques to retrieve the largest possible
information contained in the Stokes profiles. To study the effect of the
different heights of formation of the $630$ nm line pair on the
results retrieved from using these techniques, we have carried out a numerical
experiment. A synthetic data set of Stokes $I$ and $V$, covering both spectral lines, that will
be analyzed as if they were
{\it observed} profiles, has been constructed using a two component model:
a magnetic component occupying a fraction of the resolution element and a
non magnetic component filling up the rest of the space. Both components
are based on the HSRA quiet Sun model. They only differ on the 
magnetic field, which has been assumed to be
vertical and constant with height, with strengths
going from $100$ to $1000$ G, and a corresponding filling factor such that, 
in all the cases, the magnetic flux density is $10$ G (similar
to that measured in internetwork regions). No noise has been added
to the synthetic profiles to avoid any uncertainty related to this parameter. 
The reason for choosing a vertical magnetic field relies on the low linear 
polarization values that are measured. Strong fields are supposed to be predominantly 
vertical due to buoyancy effects \citep{grossmann_98} while 
weak fields are expected to have all possible inclinations  
\citep[or more horizontal,][]{sigwarth_00}. Taking 
the weak field approximation for circular and linear polarization 
(\cite{egidio}), we obtain 
a relationship between their amplitudes as follows:
\begin{equation}
\frac{\sqrt{Q^2+U^2}}{V}=0.082\frac{\sin^2 \theta}{\cos\theta} ,
\end{equation}

\noindent where $\theta$ is the inclination angle with respect to the line of sight. 
Note that, for an intermediate value ($\theta=45^o$) and typical 
circular polarization amplitudes at internetwork regions ($10^{-3} I_c$), Stokes 
$\sqrt{Q^2+U^2}$ amplitude equals $5.8\times 10^{-5} I_c$, i. e., almost a factor 
$20$ smaller than the amplitude of V. This makes them negligible and undetectable 
even with extremely low noise observations.

The synthetic set of Stokes $I$ and $V$ profiles has been inverted using the
SIR\footnote{Stokes Inversion based on Response functions} code
\citep{basilio_92}. In each inversion, the magnetic field strength has
been forced to a wrong value,
treating the rest
of the magnitudes (temperature stratification of both components,
magnetic filling factor, and micro and macroturbulent velocities) as free parameters. 
The purpose of this experiment is to find out to what extent can the effect of the magnetic
field on the Stokes profiles be compensated by the rest of the magnitudes of the model
atmospheres.

\begin{figure}[!t]
\includegraphics[width=8cm]{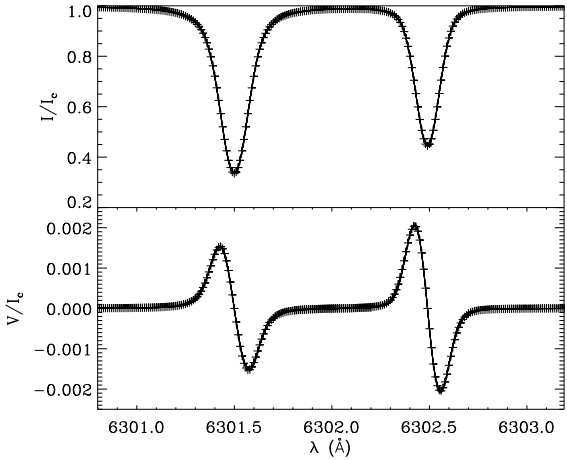}
\caption{Fits to $1000$ G synthetic $I$ and $V$ profiles at $630$ nm with different
model atmospheres having the following fixed field strengths: $100$, $300$,
$500$, $700$, $900$ and $1000$ G. Plus signs represent the synthetic profiles and lines are the fits.}
\label{ajustes}
\end{figure}

\begin{figure}[!t]
\includegraphics[width=8cm]{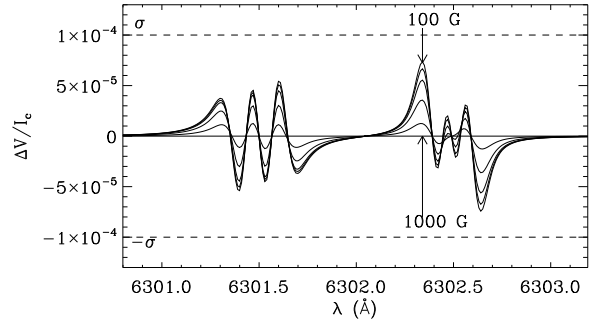}
\caption{Fitting mismatches to a $1000$ G $V$ profile with smaller field
strength model atmospheres. Dashed lines denote the typical
noise level value of present high spatial resolution spectropolarimetric
data.}
\label{iguales}
\end{figure}

Figure \ref{ajustes} shows the Stokes $I$ and $V$ synthetic profiles (plus signs) for
the case B=1000 G and, overplotted, the results of the inversion obtained
for the different fixed mistaken magnetic field strengths. The differences
between the retrieved and the input Stokes parameters are tiny in all
cases. To make it more evident, Figure \ref{iguales} displays the
difference between the retrieved Stokes $V$ profiles and the input
one. As expected, the fit with a
field strength of $1000$ G is perfect. The smaller the field strength
the higher the mismatch. If a typical value for the noise of
spectropolarimetric observations is taken to be of the order of $10^{-4}I_c$,
the discrepancies lie, in all cases, well below that noise level and very
far from the value of $3\sigma$. Therefore, we reach the conclusion that 
it is possible to reproduce the shape of the
$1000$ G $I$ and $V$ profiles with very different magnetic field strengths.

\begin{figure}[!t]
\includegraphics[width=8cm]{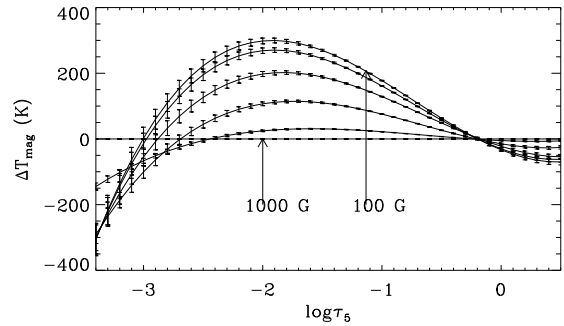}
\caption{Difference between the temperature of the initial magnetic model
atmosphere and that of the magnetic atmospheres recovered from the inversion
of a $1000$ G profile forcing the field strength to smaller values.}
\label{iguales_temp}
\end{figure}

\begin{figure}[!t]
\includegraphics[width=8cm]{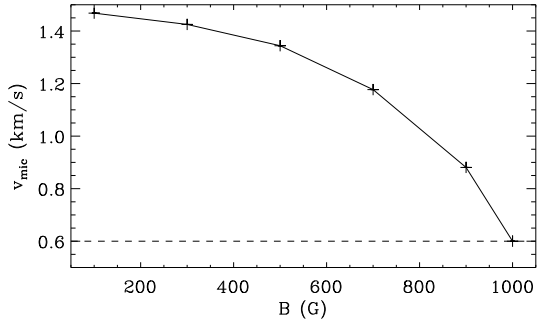}
\caption{Microturbulent velocity
resulting from the inversion of a $1000$ G profile forcing the field
strength to smaller values (solid line). The dashed line is the
initial value of the microturbulent velocity feeded to the inversion process.}
\label{iguales_mic}
\end{figure}

To confirm that these good matches represent degenerated solutions of the
problem, we have checked that the calculated temperature stratifications
are realistic. In Figure \ref{iguales_temp} we have plotted the difference
between the temperature of the ``actual'' and the retrieved magnetic component,
for each inversion with a fixed wrong value of the magnetic field. The
variations are only of a few hundred degrees in the optical depth range where the
lines are sensitive. Due to the different heights of formation of the Fe\,{\sc i} 
lines, a small change in the temperature stratification can modify the ratio of the 
amplitudes of the Stokes $V$ profiles between both lines. 
As a consequence, we are compensating the effect of the wrong magnetic field
with a slightly different temperature gradient. Figure \ref{iguales_mic}
shows the microturbulent velocity for the magnetic component recovered from the
inversions. There is a clear tendency: in order to reproduce the $1000$ G profile
with a smaller field strength, an increase of the microturbulent velocity
is needed. The excess broadening of the line is produced by a microturbulent
velocity rather than by the magnetic field.

It is important to note that, in this simple case, the modification of
the temperature gradient and the microturbulent velocity compensates
for the effect of the magnetic field strength. In more realistic situations,
magnetic field strengths and bulk velocities may vary with height.
These gradients may also affect in a different manner both spectral lines,
when taken into account in the inversion process as additional free
parameters they may introduce another source of uncertainty or
degeneracy \citep[see][]{khomenko_spw4}.

\section{The line ratio technique}

The line ratio technique, first presented by \cite{howard_72} is a
powerful method to retrieve magnetic field strengths from structures which
are not resolved. A lot of works on internetwork regions are based on this
simple method \citep{keller_94, ita_franz_03, lites_04, hector_valentin_04}, 
since it allows us to have a close idea of the magnetic
properties of these regions only taking into account the information given by Stokes V.
This method demands a pair of lines fulfilling strict requirements: the same
wavelength, line strength ($\log{gf}$), excitacion potential (same
response to the temperature), and different effective Land\'e factors.
Under these hypotheses, one can infer the magnetic field strength outside
the weak field regime. The weak field approximation can be applied to a
range of field strengths which depends on the transitions. In the case of
the $630$ nm lines, field strengths smaller 
than approximately $500$ G belong to the weak field regime. For larger field strengths,
the ratio of the amplitudes of the Stokes $V$ signals is a function of the magnetic field, 
and this makes it possible to assign a magnetic field strength given the 
amplitude ratio. To check the validity of the line ratio technique
applied to this pair of lines, we have computed the
ratio between the amplitudes of $V$ of both lines, taken
from the previously inverted profiles. As expected, since all the fits match very well,
the line ratio value ranges between $0.74 \pm 0.01$, indicating that, for this
pair of lines, the ratio between the amplitudes of Stokes $V$ is not a good indicator of
magnetic field strength with values ranging from $100$ to $1000$ G.



\section{Lack of observables}

Figure \ref{iguales} could lead to the conclusion that by decreasing the noise level to
values smaller than the mismatch, the correct magnetic field strength could be recovered.
To confirm or disregard this assumption we have performed a numerical study
synthesizing a particular Stokes vector and inverting it 
following two different schemes. 

\begin{figure}[!t]
\includegraphics[width=8cm]{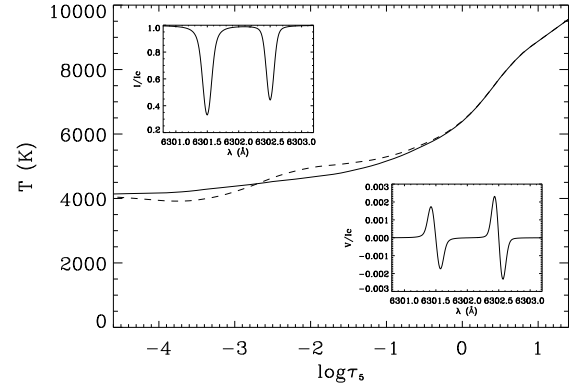}
\caption{Temperature stratifications for the synthesis of the Stokes vector used for the numerical 
test. The solid line is the temperature stratification for the non magnetic component and the dashed line 
for the magnetic one. The small plots inserted are the Stokes $I$ and $V$ profiles synthesized from them (see text for details).}
\label{dif_temp_sintesis}
\end{figure}

Direct and indirect evidences seem to indicate that the quiet 
Sun is full of magnetic elements \citep{stenflo_87, rafa_04, javier_04}. This magnetic 
background is not detectable with Zeeman diagnostic tools. Thus, the most plausible 
approximation must deal at least with two different atmospheres (magnetic and non magnetic) 
that can show a large disparity between their atmospheric variables. The two magnetic components 
of the synthesis differ on the temperature gradient, the microturbulent velocity and the 
magnetic field. The intensity and circular polarization profiles of the pair at \hbox{$630$ nm} coming from 
this two-model atmosphere are synthesized. The temperature of the magnetic component differs 
from the non magnetic one as shown in Figure \ref{dif_temp_sintesis}. The magnetic 
microturbulent velocity is set to $1.4$ km/s while the non magnetic one is fixed to $0.6$ km/s. 
The magnetic field with a strength of \hbox{$300$ G} has been supposed to be vertical and constant with 
optical depth. Consistently with the previous test, the magnetic flux density is 
forced to $10$ G in order to have 
circular polarization signals of the same order of magnitude as the ones observed. 

The synthetic $I$ and $V$ profiles with a magnetic field strength of $300$ G are inverted adding a
random noise distribution with a standard deviation of $5\times 10^{-5}I_c$ and $10^{-4}I_c$. 
For each case, one hundred realizations of the noise are done to make sure that the test 
is statistically significant. We follow two different schemes for the inversion procedure. In 
the first one, we force the temperature stratification and the microturbulence to be the same 
both in the non magnetic and in the magnetic components. And, in the second one, we allow the microturbulence 
and the temperature to vary independently at both atmospheric components. In both cases, 
the magnetic field strength and the filling 
factor are free parameters of the inversion and are initialized randomly. The rest of the variables 
are treated as in Section $3$.

\subsection{Inversion with fixed temperatures at both atmospheres}

Some works dealing with LTE inversions at internetwork regions 
\citep[see, for example,][]{hector_04} assume 
that the temperatures of both the non magnetic and the magnetic components have the same stratification 
with $\log(\tau_{5})$. The same approximation is implicit in Milne-Eddington inversions. 
The line absorption coefficient is the same for both components and it is usually fixed by the 
intensity profiles. In the same way, in this first test, the temperature stratification and 
the microturbulent velocity are forced to be the same for both components in our inversions. 
To achieve it, the same model initialization is used for the non magnetic and the magnetic 
atmosphere and the perturbations to these parameters are forced to be equal.

\begin{figure}[!t]
\includegraphics[width=8cm]{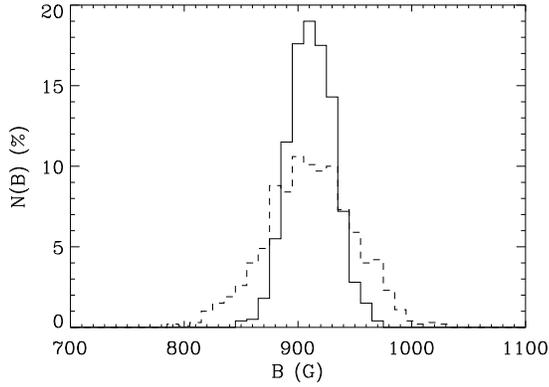}
\caption{Histogram of the inferred magnetic field strengths for the two different noise levels. The solid 
line represents the case with a noise level of $5\times 10^{-5} I_c$ and the dashed line the one with 
$10^{-4} I_c$. The inversion procedure assumes the same temperature stratification with optical 
depths and microturbulence for both magnetic and non magnetic components. The input synthetic profiles 
have been generated with a magnetic field strength of $300$ G.}
\label{igualT}
\end{figure}

Regardless of the noise level, the computed strengths concentrate 
around $900$ G. As expected, the inversions with larger noise level result in a slightly wider range 
of magnetic fields. The ``real'' temperature differences are compensated with an increase of $600$ G 
of the ``true'' magnetic field strength. The important fact is that this behaviour is independent 
of the initialization. 

\begin{figure}[!h]
\includegraphics[width=8cm]{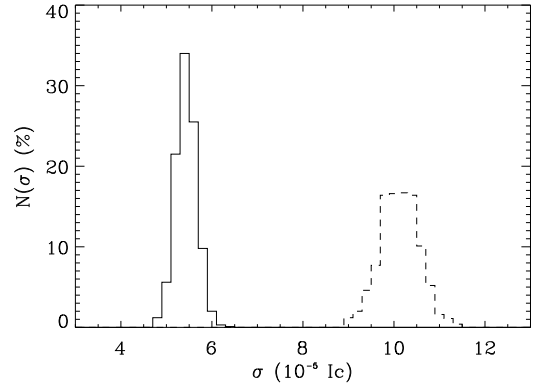}
\caption{Standard deviations of the difference between the noisy synthetic profiles and their fits. 
The solid and dashed lines represent the cases with a noise level of $5\times 10^{-5} I_c$ and $10^{-4} I_c$, respectively. 
The inversion procedure assumes the same temperature stratification with optical depths and microturbulence 
for both the magnetic and the non magnetic components.}
\label{sigma_igualT}
\end{figure}

The standard deviation histograms of the difference between the inverted and the 
synthetic $V$ profiles are plotted in Figure \ref{sigma_igualT}: all the inversions fit equally well 
the synthethic profile. The standard deviations are similar to the corresponding noise level 
in each case, meaning that all solutions to the inverse problem are acceptable. 
The assumption of two atmospheres with the same temperature 
stratification tends to overstimate the weak input magnetic field 
to almost $1$ kG in this particular case. The resulting mean magnetic flux is $11$ G, 
slightly higher than the input for the synthesis. 

\subsection{Inversion with free temperature stratifications}

In the second test, we have left the temperature and microturbulence as independent free 
parameters for both atmospheric components. If the information of the thermodynamical and magnetic properties were 
present in the Stokes $V$ profile, the inversion code should be capable of finding the 
correct solution. We have proceeded, as in the previous test, by 
initializing both components with the same model atmosphere except for the magnetic field 
strength (random initialization).

\begin{figure}[!h]
\includegraphics[width=8cm]{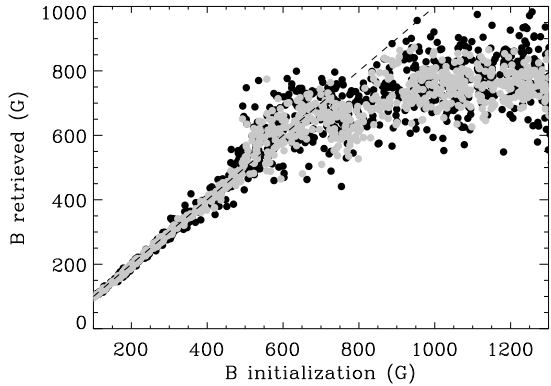}
\caption{Magnetic field strengths inferred from the inversions with the temperature stratification and the 
magnetic microturbulent velocity as free parameters. Grey dots represent the results with the noise level of $5\times10^{-5} I_c$ and the black dots the ones for $10^{-4} I_c$. The inversion procedure has the temperature stratification with 
optical depths and microturbulence as free parameters for both magnetic and non magnetic components.}
\label{difT}
\end{figure}

\begin{figure}[!h]
\includegraphics[width=8cm]{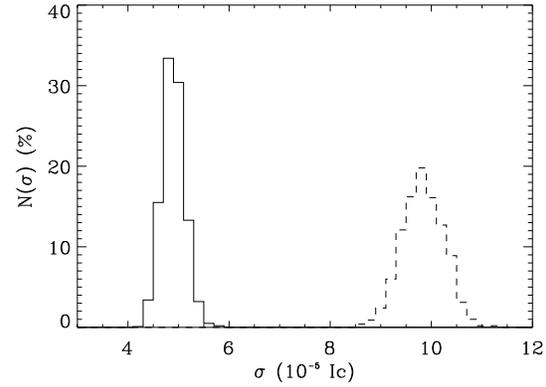}
\caption{Standard deviations of the difference between the noisy synthetic profiles and their fits.  
The solid and dashed lines represent the cases with a noise level of $5\times 10^{-5} I_c$ and $10^{-4} I_c$, respectively. The inversion procedure has the temperature stratification with 
optical depths and microturbulence as free parameters for both magnetic and non magnetic components.}
\label{sigmas_difT}
\end{figure}

\begin{figure}[!h]
\includegraphics[width=8cm]{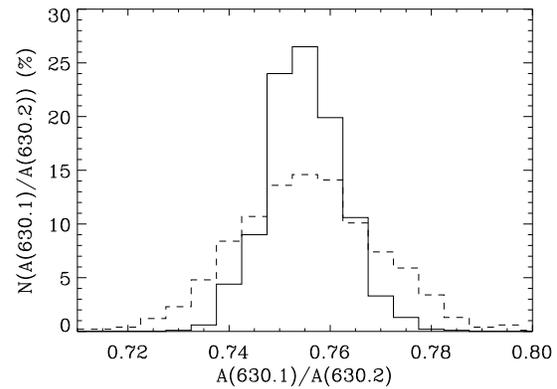}
\caption{Histogram of the ratio between the amplitudes of Stokes $V$ for the 630.1 and the 630.2 lines. The solid line represents the case with a noise level of $5\times 10^{-5} I_c$ and dashed one that of $10^{-4} I_c$. The inversion procedure sets free the temperature stratification with 
optical depths and microturbulence for both the magnetic and the non magnetic components.}
\label{ampl}
\end{figure}

The results of the inversions are presented in Figure \ref{difT}. The magnetic field strength 
inferred from the inversions is plotted as a function of the magnetic field initialization. For 
both noise realizations, while the ``true'' magnetic field is $300$ G, the computed value is 
nearly the same as the one feeded to the inversion code until $700$ G. 
Stronger field initializations return field strengths  
around $700$-$900$ G. The dependence on the input magnetic field is not a problem of the code since 
all the fits have the same quality. Figure \ref{sigmas_difT} shows the histograms of the standard 
deviation of the difference between the synthetic and the fitted Stokes $V$ profiles for each noise 
level case. Both distributions are narrow and peak at the added noise value. 
The mean magnetic flux is $10$ G. 
Figure \ref{ampl} shows the histogram of the amplitudes of the Stokes $V$ profiles with 
a narrow distribution 
around $0.755$.  For this particular 
profile, even if the noise level is as low as $5\times10^{-5} I_c$ and the temperature and 
microturbulent velocity of the magnetic component are free parameters, one can not 
distinguish between $100$ and $800$ G. Low flux internetwork Stokes spectra 
of the $630$ nm lines do not seem to carry 
enough observables to separate the thermodynamical and the magnetic effects. The 
magnetic field is compensated by temperature gradients and microturbulent velocities in the 
same way as in section $3$.


\section{Observational data}

To confirm the numerical results found in the previous paragraphs, we have
analyzed a data set corresponding to a \hbox{33" x 42"} internetwork region observed
on August 17, 2003 at the disc centre. Network regions were deliberately avoided 
using the real-time Ca K image available during the observations. The spectropolarimeter
POLIS\footnote{POlarimetric LIttrow Spectrograph} \citep{beck_05}
was used, attached to the German VTT at El Teide observatory, to obtain
full Stokes spectropolarimetric data in the 630 nm wavelength region.
The image was stabilized using a device correlation tracker \citep{ballesteros_96},
which also made possible an accurate stepping perpendicular to the slit
at steps of $0.35''$. The integration time at each slit position was around $27$
seconds. After the data reduction, a noise level of $7\times 10^{-5} I_c$ and a
spatial resolution of $1.3''$ were achieved. 

\begin{figure}[!t]
\includegraphics[width=8cm]{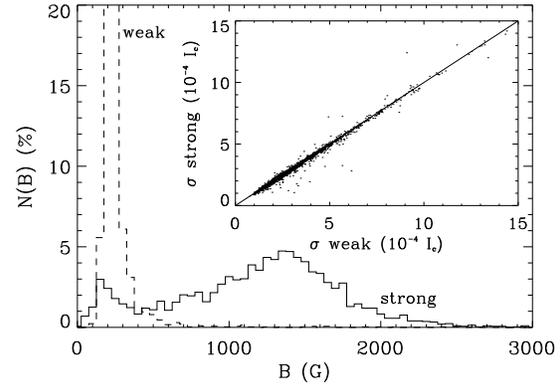}
\caption{Magnetic field distribution recovered from the observed data using two different inversions,
with a strong (solid line) and a weak (dashed line) magnetic field
strength initializations. The small plot represents the standard deviation, $\sigma$, of the difference 
between the observed and the fitted Stokes $V$ profiles. The overplotted continuum line is the 
diagonal.}
\label{chi2}
\end{figure}

\begin{figure}[!t]
\includegraphics[width=8cm]{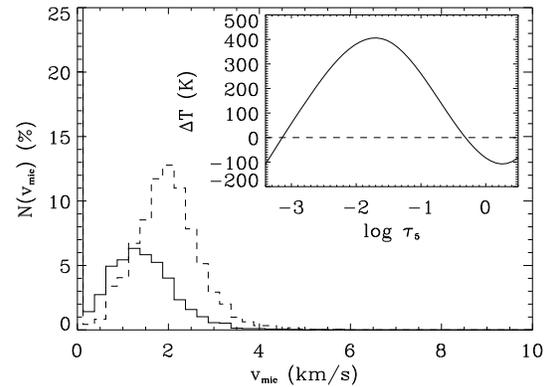}
\caption{Histograms of the microturbulent velocity recovered from two different inversions of the observed data,
with a strong (solid line) and a weak (dashed line) magnetic field 
strength initializations. The plot in the small window represents the difference between the mean temperature stratification inferred 
from the weak and the strong initialization inversions.}
\label{dif_temp_mic}
\end{figure}

As before, the SIR code was used
to invert the $630$ nm $I$ and $V$ data following two different strategies: in the first
one, we initialized the code with $1200$ G; in the second, a starting value of $200$ G was 
used. The free parameters and the model initialization were the same as in the numerical 
experiment of section 5.2. For both inversion sets, we retrieved
the same magnetic flux but very different magnetic field strength
distributions \citep{marian_spw4}. With a strong (weak) field initialization, a peak around
$1300$ ($200$) G in the PDF is obtained. Both magnetic field
distributions can be seen in Figure \ref{chi2}. The most important fact is 
that both distributions fit equally well the observations.
The small window in Figure \ref{chi2} compares the standard deviation of the 
difference between the observed and fitted Stokes $V$ profiles computed for both
initializations. Both distributions have the same $\sigma$. The conclusion is
reached that, with our experimental data, we can not distinguish
between the two magnetic field distributions presented in the figure. Temperature and microturbulent 
velocities of both components behave as predicted by the numerical tests. 
Figure \ref{dif_temp_mic} shows the histogram of the microturbulent velocity 
recovered from the inversions. Larger values are found in the weak initialization procedure 
rather than in the strong one. In the inserted graphic the difference between the 
mean temperature stratifications recovered from the weak and the strong initializations is plotted. 
Strong fields can be compensated by 
cooling the layers where the $630.1$ nm line has its maximum of sensitivity, and by heating 
the $630.2$ nm formation region. This thermal change translates in a variation 
of the number of absorbers, more for the $630.1$ nm transition and less for the $630.2$ nm line. 
Adding a higher microturbulence to broaden the line, the effect of a strong magnetic 
field can be well reproduced in terms of other atmospheric properties, resulting in an 
intrinsic degeneracy of magnetic and thermodynamic properties.


\section{Discusion and conclusions}

The work by \cite{carlos_98} suggested that the $630$ nm line pair 
can give reasonable estimates of the magnetic field vector for moderately 
unresolved weak fields. But the physical scenario they deal with differs substantially 
from that of the internetwork quiet Sun. First, the smallest filling factor used by those authors 
is $10$\% and the smallest circular polarization signal is  
$4\times 10^{-3} I_c$. These numbers are respectively $10$ and $4$ times larger than those 
characteristic of internetwork regions. And second, both the magnetic  
and the non magnetic components have the 
same temperature stratification for their synthetic data and in the model atmospheres 
resulting from their inversions. Thus, they do not study the effect of different thermodynamical 
(magnetic and non magnetic) properties. Their results are compatible with the assumptions 
made during their tests. As presented in this work, the relaxation of the constraint of equal 
temperature stratifications leads to different conclusions.

In this paper, a set of arguments have been presented that prove that the Fe\,{\sc i} line pair
at 630 nm is not reliable for the magnetic field strength determination in weak
flux quiet Sun regions. The reason lies in the separate heights of formation 
of both lines, that makes them sensitive to different layers in the solar
atmosphere. Model atmospheres with different magnitudes and gradients
can result in undistinguishable Stokes profiles, with differences
well below typical noise values of present high spatial resolution
spectropolarimetric observations. Slightly different temperature stratifications, combined with
adequate microturbulent velocities and magnetic field strengths, ranging
from a few hundred of Gauss to kiloGauss, give rise to the same Stokes profiles.

Previous estimations of magnetic field strengths at internetwork quiet Sun regions based on the 
630 nm lines assume, explicitly, that the temperature stratification of the magnetic element is 
like the one of the non magnetic quiet Sun. The present paper reveals that the inferred magnetic 
properties of the internetwork are strongly coupled to the hypotheses made to constrain the 
analysis. Thus, those cited works must be regarded with care until the knowledge of the 
internetwork atmosphere is precise enough as to confirm their hypotheses. 

In addition, velocity and magnetic field gradients (not considered in this work) may
lead to a larger degeneracy of the problem when using these lines. This
suggests that the possibility of using alternative lines for the study of the quiet Sun 
magnetism, such as the Fe\,{\sc i} pair at 525 nm, or the mentioned $630$ nm line pair 
together with other spectral lines to constrain the problem, needs to be explored.

\begin{acknowledgements}

Many thanks are due to Arturo L\'opez Ariste and Andr\'es Asensio Ramos for helpful 
discussions, to H\'ector Socas Navarro for sharing his view on the subject of this paper,
to Rebecca Centeno and Elena Khomenko for carefully reading the manuscript  
and to an anonymous referee who helped to improve it.
This research has been funded by the
Spanish Ministerio de Educaci\'on y Ciencia
through project AYA2004-05792.
\end{acknowledgements}





\end{document}